\documentclass[conference]{IEEEtran}
\IEEEoverridecommandlockouts
\usepackage{cite}
\usepackage{amsmath,amssymb,amsfonts}
\usepackage{algorithmic}
\usepackage{graphicx}
\usepackage{textcomp}
\usepackage{xcolor}
\usepackage{booktabs}
\usepackage{algorithm}
\usepackage{tikz}
\usetikzlibrary{arrows.meta,positioning,shapes.misc}
\usepackage[inline]{enumitem}
\usepackage{float}
\usepackage{placeins}
\usepackage{subfig}
\usepackage{hyperref}
\hypersetup{
    colorlinks=true,
    linkcolor=blue,
    citecolor=blue,
    urlcolor=blue
}
\def\BibTeX{{\rm B\kern-.05em{\sc i\kern-.025em b}\kern-.08em
    T\kern-.1667em\lower.7ex\hbox{E}\kern-.125emX}}
\begin{document}

\title{Benchmarking Early Deterioration Prediction Across Hospital-Rich and MCI-Like Emergency Triage Under Constrained Sensing 
}

\author{\IEEEauthorblockN{KMA Solaiman}
\IEEEauthorblockA{\textit{Dept. of CSEE} \\
\textit{Uni. of Maryland Baltimore County}\\
Maryland, USA \\
ksolaima@umbc.edu}
\and
\IEEEauthorblockN{Joshua Sebastian}
\IEEEauthorblockA{\textit{Dept. of CSEE} \\
\textit{Uni. of Maryland Baltimore County}\\
Maryland, USA \\
cj48611@umbc.edu}
\and
\IEEEauthorblockN{Karma Tobden}
\IEEEauthorblockA{\textit{Dept. of CSEE} \\
\textit{Uni. of Maryland Baltimore County}\\
Maryland, USA \\
ktobden1@umbc.edu}
}

\maketitle




\begin{abstract}
Emergency triage decisions are made under severe information constraints, yet most data-driven deterioration models are evaluated using signals unavailable during initial assessment. We present a leakage-aware benchmarking framework for early deterioration prediction that evaluates model performance under realistic, time-limited sensing conditions. Using a patient-deduplicated cohort derived from MIMIC-IV-ED, we compare hospital-rich triage with a vitals-only, MCI-like setting, restricting inputs to information available within the first hour of presentation. Across multiple modeling approaches, predictive performance declines only modestly when limited to vitals, indicating that early physiological measurements retain substantial clinical signal. Structured ablation and interpretability analyses identify respiratory and oxygenation measures as the most influential contributors to early risk stratification, with models exhibiting stable, graceful degradation as sensing is reduced. This work provides a clinically grounded benchmark to support the evaluation and design of deployable triage decision-support systems in resource-constrained settings.
\end{abstract}


\section{Introduction}
\label{sec:introduction}

Early identification of clinical deterioration in the emergency department (ED) is critical for timely escalation of care and resource allocation. Patients who require unanticipated intensive care unit (ICU) admission or who experience in-hospital mortality often exhibit physiological instability early in their ED course, yet accurately identifying such risk remains challenging during initial triage~\cite{salt2008, royal2017news2}. Decision-making at this stage is constrained by limited information, time pressure, and incomplete measurements -- conditions that are even more pronounced in mass-casualty--like (MCI-like) emergency response settings~\cite{sacco2005saves, xu2023mci}. 

Recent advances in machine learning have shown promise for ED risk prediction~\cite{chang2024edtriage, liu2025gradient, sitthiprawat2025electronic}, but most existing models are developed under hospital-rich assumptions, relying on laboratory results, longitudinal observations, or post-triage documentation that may not be available at the time of initial assessment~\cite{roberts2023mimiced} in MCI-like environment. Moreover, evaluation practices frequently suffer from patient-level leakage, repeated-visit contamination, or the inclusion of downstream features that would not be available to clinicians at decision time~\cite{roberts2023mimiced}. These issues limit the clinical interpretability, deployability, and comparability of prior work -- particularly for settings characterized by partial observability and constrained sensing.

In parallel, there is growing interest in understanding which physiological signals are essential for early risk stratification, rather than simply maximizing predictive performance under ideal data availability~\cite{start1996, bmj2021news2}. For triage applications, robustness to missing measurements, redundancy among vital signs, and interpretability of model behavior are as important as headline accuracy~\cite{lundberg2017shap}. However, systematic benchmarks that explicitly contrast hospital-rich and field-constrained regimes, while enforcing strict leakage control and patient-level independence, remain scarce~\cite{roberts2023mimiced}.

In this work, we address these gaps by framing early ED deterioration prediction as a decision problem under partial observability. Using a large, curated cohort derived from MIMIC-IV-ED~\cite{johnson2023mimiciv}, we construct a leakage-aware benchmarking framework that restricts features to those available within the first hour of ED arrival and enforces patient-level independence across all evaluations. We explicitly contrast a hospital-rich regime with a vitals-based, MCI-like regime to quantify the performance gap induced by partial observability and constrained sensing, consistent with prior work on major-incident triage~\cite{sacco2005saves, xu2023mci}. Beyond analysis, our framework defines a reusable benchmark that enables controlled, leakage-aware evaluation of triage models under progressively constrained sensing regimes, supporting reproducible comparison across realistic deployment settings. Through structured vital-sign ablations and interpretability analyses using SHAP~\cite{lundberg2017shap}, we further examine which physiological signals are indispensable for early risk prediction and how models degrade under reduced sensing -- an increasingly relevant concern as interest grows in deployable, sensor-limited triage systems. 

\textit{Contributions.} This work makes the following contributions:





\begin{enumerate}[leftmargin=3pt]
    \item \textbf{Problem formulation under realistic information constraints.}  
    We frame early emergency department deterioration prediction as a decision problem under partial observability, explicitly distinguishing hospital-rich triage from vitals-based, MCI-like settings using only information available during initial assessment.

    \item \textbf{Leakage-aware, reproducible triage benchmarking framework.}  
    We introduce a patient-level, reproducible benchmarking framework derived from MIMIC-IV-ED that enforces strict encounter independence, excludes post-decision information, and supports controlled evaluation across progressively constrained sensing regimes.

    \item \textbf{Physiologically grounded insight into deployable triage signals.}  
    Through systematic vital-sign ablations and interpretability analysis, we identify respiratory and oxygenation measures as the dominant contributors to early risk prediction and demonstrate robust, graceful degradation under reduced sensing.

    \item \textbf{Interpretable baselines for constrained triage modeling.}  
    We establish strong, interpretable baselines spanning linear, ensemble, and modern deep tabular models, for early deterioration prediction across both hospital-rich and MCI-like regimes, providing a foundation for future comparison.
\end{enumerate}

\section{Related Work}
\label{sec:related-v5}

Emergency triage has traditionally relied on rule-based scoring systems designed for rapid prioritization. Early warning scores such as NEWS2 apply fixed thresholds over a small set of vital signs and remain widely deployed in hospital settings, though performance varies across populations and clinical contexts \cite{royal2017news2,bmj2021news2}. In mass-casualty and prehospital environments, lightweight protocols such as START, SALT, and AVPU-based assessments enable decision-making under severe time and resource constraints \cite{start1996,sacco2005saves,salt2008}. While operationally practical, these approaches provide limited insight into how individual physiological signals relate to downstream clinical deterioration. Clinical studies show that initial vital signs—including blood pressure, heart rate, oxygen saturation, and level of consciousness—are associated with short-term mortality, though individual measures exhibit modest discriminative power in isolation \cite{erwander2025vitals}.

Learning-based methods have demonstrated improved performance over traditional triage scores for predicting adverse outcomes in emergency care. Interpretable models such as logistic regression remain competitive baselines \cite{boulitsakis2023predicting}, while ensemble methods and gradient boosting achieve strong discrimination in hospital-rich settings \cite{sitthiprawat2025electronic}. Neural models incorporating free-text triage notes further improve predictive accuracy \cite{chang2024edtriage}, and interpretable deep learning systems have been proposed for predicting triage-related outcomes such as severity level, hospitalization, and length of stay \cite{lin2024interpretable}. In major-incident settings, machine learning–derived triage tools based on trauma registries have been explored to improve sensitivity for high-acuity patients \cite{xu2023mci}. However, most prior work relies on hospital-available information, including laboratory results or longitudinal observations, and does not assess robustness when prediction is restricted to the limited vital signs available at initial assessment.

Public datasets such as MIMIC-IV and MIMIC-IV-ED have enabled reproducible research on emergency and inpatient prediction tasks \cite{johnson2023mimiciv,johnson2024mimic}. Despite their widespread adoption, these resources lack standardized benchmarks tailored to early triage under realistic information constraints. Recent efforts such as \cite{PhysioNet-mietic-1.0.0} repurpose MIMIC-IV-ED to model Emergency Severity Index (ESI) assignments at intake, focusing on reproducing clinician triage decisions rather than predicting downstream clinical deterioration under constrained sensing. Many studies rely on site-specific cohorts or heterogeneous feature definitions, limiting comparability across methods and care settings \cite{roberts2023mimiced}, while trauma registries used in mass-casualty research are typically not openly accessible \cite{xu2023mci}. As a result, infrastructure for systematically contrasting hospital-rich and field-constrained triage regimes within a unified, leakage-aware evaluation framework remains limited.

Interpretability is widely recognized as essential for the adoption of predictive models in emergency triage  \cite{liu2021prospective}. SHAP is commonly used to explain tabular clinical prediction models, including ICU deterioration and mass-casualty triage tasks \cite{lundberg2017shap,liu2025gradient,xu2023mci}, while alternative explanation methods have been explored in text-rich settings \cite{chang2024edtriage}. However, few studies link interpretability to robustness under constrained sensing or to comparative evaluation across distinct triage regimes.

Our work addresses these gaps by framing early deterioration prediction as a benchmarking problem under constrained sensing. We introduce a leakage-aware, reproducible evaluation framework derived from MIMIC-IV-ED that contrasts hospital-rich and MCI-like triage regimes using only information available at initial assessment. Beyond comparing predictive performance, we examine how model accuracy degrades as sensing is restricted and identify which physiological signals remain essential for early risk stratification, providing interpretable baselines to guide deployable decision-support systems.

\section{Dataset Construction}
\label{sec:dataset}

\subsection{Data Source}
We derive the cohort from the publicly available \textit{Medical Information Mart for Intensive Care IV} (MIMIC-IV v3.1) \cite{johnson2023mimiciv} and its Emergency Department module (MIMIC-IV-ED v2.2) \cite{johnson2024mimic}. We begin from emergency department (ED) encounters recorded in \texttt{edstays.csv} and link them to inpatient admissions (\texttt{admissions.csv}), ICU stays (\texttt{icustays.csv}), patient demographics (\texttt{patients.csv}), ED vital signs (\texttt{vitalsign.csv}), triage forms (\texttt{triage.csv}), and laboratory events (\texttt{labevents.csv}). A limited, prefiltered subset of \texttt{chartevents.csv} is incorporated to obtain early clinical observations, including consciousness proxies (e.g., GCS verbal response) and respiratory assessments, when available.

All data are extracted to reflect information available during the initial ED assessment window, prior to downstream inpatient or ICU interventions. Record linkage uses clinically meaningful keys to prevent cross-admission leakage: \texttt{edstays} is linked to \texttt{admissions} via (\texttt{subject\_id}, \texttt{hadm\_id}); patient demographics are linked via \texttt{subject\_id}; vital signs and triage forms via (\texttt{subject\_id}, \texttt{stay\_id}); laboratory events via \texttt{hadm\_id}; and ICU stays via (\texttt{subject\_id}, \texttt{hadm\_id}), retaining only the first ICU \texttt{intime} per admission.

MIMIC-IV is fully de-identified and accessible to credentialed researchers through PhysioNet, enabling reproducible research. To support scalable processing of the full cohort, data are loaded in chunks and filtered during ingestion, enabling efficient construction on modest hardware.

\subsection{Cohort Inclusion/Exclusion and Data Normalization}

We restrict the cohort to adult patients ($\geq$18 years) with a documented emergency department (ED) triage episode and at least one complete set of ED vital signs. To ensure patient-level independence, only the first ED encounter per patient is retained, also reflecting MCI-like scenarios in which only a single initial assessment is available per patient.

Vital signs and laboratory measurements are restricted to the first-hour window following ED arrival, reflecting information available during initial triage and prior to downstream clinical interventions. Physiologically implausible values (e.g., heart rate $<20$ or $>250$ bpm) are removed using clinically motivated rule-based ranges.

Continuous features are standardized using z-score normalization, with scaling parameters estimated exclusively on the training split and applied unchanged to validation and test splits. Categorical features are encoded using consistent, low-cardinality vocabularies with an explicit \texttt{unknown} category. Missing values are imputed using the training-set mean for continuous variables and the \texttt{unknown} category for categorical variables. All imputation and normalization parameters are learned on training folds only to prevent information leakage.


\subsection{Outcome Definition}

We formulate early deterioration prediction as a binary classification task. The \emph{primary outcome} is in-hospital mortality during the same admission following emergency department presentation (label = 1). All surviving patients are assigned the negative class (label = 0).
To examine endpoint-specific sensitivity, we additionally evaluate \emph{ICU transfer within 24 hours of ED arrival} as a \emph{secondary deterioration outcome}. ICU transfer reflects short-horizon escalation driven by acute physiological instability, in contrast to mortality, which captures broader systemic severity over the course of hospitalization.
Unless otherwise stated, all main results, ablation analyses, and interpretability findings focus on the mortality endpoint. ICU transfer results are reported separately to contrast feature importance and model behavior across deterioration definitions.


\subsection{Feature Regimes}
To study early deterioration prediction under varying information availability, we define feature regimes reflecting realistic triage contexts.
\begin{itemize}
    \item \textbf{Hospital-rich regime.}
The hospital-rich regime includes all features available during the initial ED assessment window, comprising triage-time vital signs, structured clinical observations, laboratory measurements, and limited early charted indicators. This setting represents a well-resourced hospital environment in which downstream diagnostic information is accessible shortly after presentation.

\item \textbf{MCI-like regime.}
The MCI-like regime restricts inputs to basic physiological vital signs recorded at triage, including heart rate, respiratory rate, systolic blood pressure, temperature, and oxygen saturation. 
AVPU indicators and respiratory/oxygenation flags (reflecting bedside physiological assessment at initial triage) are treated as part of the vitals feature set, consistent with prior ED deterioration studies \cite{boulitsakis2023predicting}. This configuration reflects  organized mass-casualty and emergency response settings characterized by partial observability, limited sensing, and constrained resources, where laboratory testing and detailed clinical documentation may be unavailable. 

\item \textbf{Intermediate feature availability.}
To better understand how additional information contributes beyond vitals alone, we additionally analyze intermediate feature sets that incorporate structured triage observations and labs. These analyses are used to quantify incremental gains from observations but are not treated as the primary MCI-like regime. 
\end{itemize}

Unless otherwise noted, baseline performance comparisons focus on the hospital-rich and vitals-only regimes, while ablation and cumulative feature analyses characterize robustness under partial observability.
Table~\ref{tab:feature-groups} lists the final variables exposed in each regime.


\begin{table}[h]
\centering
\caption{Feature groups by regime. (\checkmark: included; --: excluded)}
\label{tab:feature-groups}
\footnotesize
\setlength{\tabcolsep}{4pt}
\begin{tabular}{@{}lcc@{}}
\toprule
Feature group & Hospital-rich & MCI-like (vitals-only) \\
\midrule
Demographics (age, sex) & \checkmark & -- \\
Vitals (T, HR, RR, SBP/DBP, SpO$_2$) & \checkmark & \checkmark \\
AVPU status & \checkmark & \checkmark \\
Derived resp./oxygen flags & \checkmark & \checkmark \\
Triage observations (pain, acuity) & \checkmark & -- \\
Chief complaint indicators & \checkmark & -- \\
Race/ethnicity (bag-of-words) & \checkmark & -- \\
Early labs (Hb, BUN, Na, K, Cr) & \checkmark & -- \\
Service utilization proxies & \checkmark & -- \\
\bottomrule
\end{tabular}
\end{table}

\subsection{Stepwise Construction and Record Linkage}
We implement a deterministic pipeline to construct a reproducible, leakage-safe triage cohort aligned to the initial ED assessment window --

\begin{enumerate}
    \item \textbf{Load \& time-align}: parse timestamps; compute ED arrival; retain only measurements within the first-hour window relative to ED arrival. To reduce memory footprint on large tables (e.g., \texttt{chartevents}, \texttt{labevents}), we first identify target subject IDs from a prefiltered index and load data in chunks, filtering during read rather than after—enabling processing of the full MIMIC-IV cohort on modest hardware.

    \item \textbf{Initial vitals}: sort \texttt{vitalsign} by (\texttt{subject\_id}, \texttt{stay\_id}, \texttt{charttime}); retain the first complete vital sign set per ED stay.

    \item \textbf{Admissions/ICU linkage}: join \texttt{edstays} to \texttt{admissions} via (\texttt{subject\_id}, \texttt{hadm\_id}); link to the first ICU stay if present; drop duplicate merges (\texttt{\_x}/\texttt{\_y}) with precedence for ED-timestamped values.

    \item \textbf{Labs}: map test labels in \texttt{d\_labitems} to canonical IDs; filter \texttt{labevents} to target tests; take the earliest value per (\texttt{hadm\_id}, test) within one hour of arrival; pivot to wide format.

    \item \textbf{Structured observations}: merge \texttt{triage} and \texttt{vitalsign} on (\texttt{subject\_id}, \texttt{stay\_id}); normalize categorical fields (e.g., pain, acuity); tokenize chief complaint for optional bag-of-words indicators. These features are used in the hospital-rich regime and in intermediate analyses of incremental feature availability beyond vitals.

    \item \textbf{Consciousness and respiratory proxies}: from a filtered subset of \texttt{chartevents}, extract early indicators including GCS verbal response (\texttt{itemid} 223900), respiratory device indicators (28 \texttt{itemids}), oxygenation parameters (6 \texttt{itemids}), and nausea/vomiting assessments (4 \texttt{itemids}). For each proxy type, we retain the earliest observation within the one-hour window.
\end{enumerate}

\paragraph*{Derived features}
We add the following pre-computed clinically interpretable features to the original cohort to support robust modeling under partial observability:
\begin{enumerate}
    \item \textbf{AVPU status}: GCS verbal responses are mapped to AVPU categories (A, V, P, U) with one-hot encoding and an explicit \texttt{unknown} category. AVPU indicators are treated as part of the vitals feature set.

    \item \textbf{Respiratory and oxygenation status}: heterogeneous respiratory observations are harmonized into compact indicators, including a binary \texttt{resp\_abnormal} flag and oxygen support variables. These derived proxies are included in the vitals feature set for both regimes.

    \item \textbf{Oxygen support and ventilation}: device mentions are normalized to clinically meaningful categories (e.g., room air, low-flow cannula, NIV, invasive ventilation) with a binary \texttt{on\_oxygen} flag. Oxygenation parameters capture FiO$_2$, O$_2$ flow rate, and delivery device configuration.  When direct FiO$_2$ measurements are unavailable, we estimate FiO$_2$ from oxygen flow rates using FiO$_2$ \(\approx 0.21 + 0.04 \times\) flow rate, capped at 1.0.

    \item \textbf{Optional symptom-text features}: select nursing assessment text fields (e.g., nausea/vomiting descriptors) are encoded using a clinical BERT model with PCA reduction for compatibility with tree-based models. 
    We report modeling primarily with structured vitals and proxies; text-derived components are included only when explicitly enabled.


    \item \textbf{Shock index}: HR/SBP computed from first-hour vitals.

    \item \textbf{Demographics}: age, sex, race/ethnicity, triage acuity, pain score.
    \item \textbf{Service utilization:} same-day emergency care (SDEC), and 30-day readmission flags.
\end{enumerate}

\begin{figure}[t]
\centering
\resizebox{0.95\columnwidth}{!}{%
\begin{tikzpicture}[
  node distance=6mm and 6mm,
  box/.style={draw, rounded corners, align=left, inner sep=3pt, fill=gray!7},
  callout/.style={draw, rounded corners, align=left, inner sep=3pt, fill=blue!6},
  opt/.style={draw, rounded corners, align=left, inner sep=3pt, fill=green!8},
  arrow/.style={-{Stealth[length=2mm]}, thick}
]

\node[box] (load) {Load \& filter by subject\\ \scriptsize Memory-efficient loading; filter during read};
\node[box, below=of load] (vitals) {Initial vitals (readmission-aware)\\ \scriptsize First complete set per (subject\_id, stay\_id)};
\node[box, below=of vitals] (link) {Admissions/ICU linkage\\ \scriptsize edstays$\leftrightarrow$admissions (subject\_id, hadm\_id); first ICU intime};
\node[box, below=of link] (labs) {Labs (first-hour)\\ \scriptsize Map d\_labitems; Hb, BUN, Na, K, Cr; earliest per hadm\_id};
\node[box, below=of labs] (obs) {Observations \& notes\\ \scriptsize Triage joins; race BoW; complaint flags; proxies};

\node[opt, below=of obs] (bert) {BERT embeddings\\ \scriptsize Clinical Sentence-BERT for nausea/vomiting; PCA to 384d};

\node[box, below=of bert] (resp) {Consciousness/respiratory/FiO$_2$ proxies\\ \scriptsize GCS verbal$\rightarrow$AVPU (one-hot); breathing (9-cat); FiO$_2$};
\node[opt, below=of resp] (svc) {Service utilization\\ \scriptsize SDEC; 30-day readmission; disposition};
\node[box, below=of svc] (label) {Label\\ \scriptsize In-hospital mortality (or ICU$<24$h)};
\node[box, below=of label] (clean) {Clean \& normalize\\ \scriptsize Outliers, imputation, scaling};
\node[box, below=of clean] (reg) {Regimes \& export\\ \scriptsize Hospital-rich vs.\ MCI-like};

\draw[arrow] (load) -- (vitals);
\draw[arrow] (vitals) -- (link);
\draw[arrow] (link) -- (labs);
\draw[arrow] (labs) -- (obs);
\draw[arrow] (obs) -- (bert);
\draw[arrow] (bert) -- (resp);
\draw[arrow] (resp) -- (svc);
\draw[arrow] (svc) -- (label);
\draw[arrow] (label) -- (clean);
\draw[arrow] (clean) -- (reg);

\node[callout, right=10mm of vitals] (c1) {\scriptsize \textbf{Process:}\\ groupby-first recipe; repeat-encounter handling};
\node[callout, right=10mm of link] (c2) {\scriptsize \textbf{Process:}\\ join keys \& dedup policy};
\node[callout, right=10mm of labs] (c3) {\scriptsize \textbf{Process:}\\ itemid mapping; earliest-value rule};
\node[callout, right=10mm of obs] (c4) {\scriptsize \textbf{Process:}\\ race normalization; complaint parsing};
\node[callout, right=10mm of resp] (c5) {\scriptsize \textbf{Process:}\\ AVPU mapping; breathing 9-cat; FiO$_2$ estimation};

\draw[arrow] (vitals.east) -- (c1.west);
\draw[arrow] (link.east) -- (c2.west);
\draw[arrow] (labs.east) -- (c3.west);
\draw[arrow] (obs.east) -- (c4.west);
\draw[arrow] (resp.east) -- (c5.west);

\end{tikzpicture}%
}
\caption{Overview of the deterministic cohort construction and preprocessing pipeline. The workflow enforces first-hour temporal alignment, patient-level deduplication, and leakage-safe record linkage, and produces hospital-rich and MCI-like feature regimes for downstream benchmarking.}
\label{fig:full_pipeline}
\end{figure}
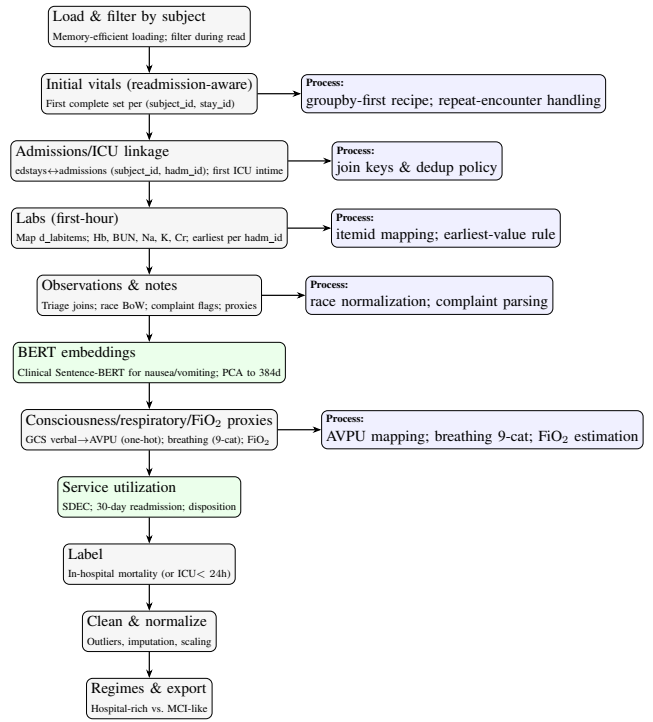

Fig.~\ref{fig:full_pipeline} shows the full dataset construction pipeline. 
Feature dictionaries and mapping specifications (e.g., AVPU derivation from GCS verbal responses, respiratory and oxygenation normalization, laboratory item mappings) are documented as part of the pipeline and applied consistently across experiments. Intermediate representations are stored in Apache Feather format to enable efficient columnar access and scalable preprocessing.


\subsection{Planned Artifact} 
We provide a fully specified cohort construction and preprocessing pipeline, including scripts for record linkage, time alignment, feature normalization, and regime construction (hospital-rich and MCI-like). Fixed train/test splits and baseline model configurations are defined to ensure reproducibility of all reported results.

All experimental results in this paper are computed on a curated cohort of approximately 10{,}000 emergency department encounters derived from MIMIC-IV-ED using the deterministic construction pipeline described in Section~\ref{sec:dataset}. This cohort is patient-level deduplicated and reflects only information available during the initial ED assessment window. Table \ref{tab:dataset-stats} report statistics on the current cohort.
Importantly, the cohort definition is not fixed to the demo subset. The same deterministic scripts apply directly to the full MIMIC-IV-ED cohort (approximately 425{,}000 ED stays), allowing researchers to scale experiments by pointing the pipeline to larger data extracts.

We plan to release expanded benchmark results, documentation, and standardized evaluation protocols in future work, following PhysioNet-compliant distribution practices.


\begin{table}[t]
\centering
\caption{Cohort summary statistics for the 10k-patient MIMIC-IV-ED cohort used in this paper.}
\label{tab:dataset-stats}
\footnotesize
\setlength{\tabcolsep}{2pt}
\begin{tabular}{lcc}
\toprule
\textbf{Metric} & \textbf{Value} & \textbf{Notes} \\
\midrule
Unique patients & 10{,}000 & adults ($\ge$18y) \\
ED encounters & 10{,}000 & first ED encounter/patient \\
Critical event ($y{=}1$) & 1{,}211 (12.10\%) & in-hospital mortality \\
Secondary Outcome & 4{,}514 (45.10\%) & ICU $ < $ 24hrs \\
\midrule
Age (mean $\pm$ SD) & 62.5 $\pm$ 17.4 & years \\
Male / Female & 54.6\% / 45.4\% & of cohort \\
\midrule
Temp missing / HR missing & 8.1\% / 5.9\% & first-hour window \\
RR missing / SBP missing & 6.0\% / 5.9\% & first-hour window \\
SpO$_2$ missing & 6.7\% & first-hour window \\
\bottomrule
\end{tabular}
\end{table}

\section{Problem Definition and Experimental Setup}

\subsection{Prediction Task and Outcomes}
\label{sec:task}

We study early emergency triage risk prediction under strict information constraints. The task is framed as a binary classification problem: given data available within the first hour of emergency department (ED) presentation, predict whether a patient will experience clinical deterioration.


Our primary outcome is in-hospital mortality, predicted using information available within the first hour of emergency department (ED) presentation. Mortality is chosen as a clinically meaningful indicator of severe deterioration that is routinely documented and consistently available across encounters.

In addition to the primary mortality outcome, we evaluate unanticipated ICU transfer within 24 hours of ED arrival as a secondary outcome. This auxiliary analysis is used to contrast early escalation decisions with longer-horizon deterioration risk, and to examine whether different physiological signals are emphasized under these two labeling schemes.

Critically, all features are restricted to those observable during the initial ED assessment window. No longitudinal trajectories, downstream interventions, or information from subsequent encounters are used. This design choice serves two purposes: (i) it prevents leakage from post-triage clinical actions, and (ii) it mirrors operational constraints in early emergency response and mass-casualty–like (MCI-like) scenarios, where decisions must be made under partial observability and severe time pressure. 

\paragraph*{Feature Regimes and Triage Settings}
\label{sec:feature-regimes}

To evaluate triage prediction under different levels of information availability, we consider two feature regimes derived from the same cohort.
In the \textbf{hospital-rich} regime, models use all features available within the first hour of ED presentation, including vital signs, triage observations, early laboratory results, and derived clinical indicators.
In the \textbf{MCI-like} regime, inputs are restricted to physiological measurements and assessment signals plausibly obtainable at the point of first contact, including core vital signs, AVPU consciousness states and respiratory/oxygenation flags, which are treated as part of the vitals feature set, while laboratory measurements and downstream utilization variables are excluded.

\subsection{Models}
\label{sec:models}

We evaluate a set of baseline models commonly used for tabular clinical prediction, including linear, ensemble-based, and neural approaches. All models operate on static feature snapshots constructed from information available within the first hour of ED presentation.
Model selection is guided by two considerations: (i) suitability for snapshot-based prediction using early triage information, and (ii) interpretability, which is critical in clinical decision-support settings.

Specifically, we consider Logistic Regression (LR), Random Forest (RF), XGBoost (XGB), and LightGBM (LGBM) as representative linear and tree-based baselines. In addition, we include TabNet \cite{arik2021tabnet} as a modern neural architecture for tabular data. TabNet employs sequential attention-based feature selection, enabling instance-level interpretability while operating on static inputs.

We intentionally do not consider temporal or recurrent models. \textit{While sequence-based architectures are commonly used for deterioration prediction, they rely on longitudinal measurements that are explicitly excluded in our setting.} 
TabNet allows us to assess whether neural tabular models offer advantages over tree-based ensembles without introducing temporal dependencies.

\subsection{Training and Evaluation Protocol}
\label{sec:train-eval}

All experiments use patient-level stratified train/validation/test splits (63/7/30), constructed via a sequential holdout procedure to prevent repeated-visit leakage. For robustness analyses, results are averaged over five independent splits with different random seeds, and we report \textbf{mean $\pm$ SD} (standard deviation) for threshold-independent metrics, \textit{AUPRC and AUROC}. Class imbalance is addressed through stratification and class weighting where appropriate.
To emphasize robustness and reproducibility, we train all models using fixed default hyperparameter values, rather than extensive tuning. 

Preprocessing steps, including imputation, scaling, and categorical encoding, are performed within the training portion of each split and applied to validation and test sets to avoid information leakage. 


For model performance, our primary analyses rely on AUROC and AUPRC, which capture ranking quality and are more informative under the class imbalance typical of early deterioration prediction. 
%
\textit{Threshold-dependent metrics, including F1-score and accuracy}, are computed at operating points selected on validation data for each feature regime and reported for representative splits to illustrate operational behavior. Because these metrics depend on threshold choice and class prevalence, they are not used for primary model comparison. 

Unless otherwise noted, all metrics are computed on patient-level held-out test sets to ensure independence between training and evaluation data.

\subsection{Text Feature Encoding}
\label{sec:text-features}
A limited subset of \texttt{notes} related to \textit{nausea and vomiting} is incorporated as an auxiliary signal when available. Specifically, selected charted text fields (e.g., emesis appearance and output; CIWA/CINA-related items) are encoded using a pretrained Biomedical BERT model, BiomedBERT \cite{pubmedbert}.

Sentence embeddings are extracted at the item level and subsequently reduced via principal component analysis (PCA) to 384 dimensions per item to control feature dimensionality and mitigate overfitting in the low-sample regime. PCA is fit on the full corpus of the corresponding text feature and applied unchanged to train, validation and test sets.
These text-derived features are included only in the hospital-rich regime and are not assumed to be available in field or MCI-like settings, hence this design choice does not affect label leakage or the primary conclusions of the study.

\subsection{Vital-Sign Ablation Protocol}
\label{sec:ablation}

To assess robustness under partial observability and to quantify the relative importance of physiological signals, we conduct a structured vital-sign ablation study. This protocol is motivated by field triage and mass-casualty–like (MCI-like) scenarios, where certain measurements may be unavailable due to time pressure, sensor failure, or resource constraints.

Starting from a vitals-only configuration, we evaluate three ablation strategies: (i) comparison between hospital-rich and vitals-only regimes, (ii) leave-one-out ablation, where individual vital signs are removed from the vitals set (e.g., SpO2, systolic blood pressure,
respiratory rate, temperature, heart rate), and (iii) progressive reduction to subsets containing two or one vital sign. Models are retrained and evaluated under each ablation condition using identical data splits and training procedures.

All experiments are conducted on the 10k-patient cohort using five repeated patient-level stratified train/validation/test splits with different random seeds, and results are reported as \textbf{mean ± SD} of test AUROC and AUPRC.
Performance differences across ablations are used to characterize signal importance and resilience to incomplete sensing, rather than to optimize performance.

\subsection{Interpretability Analysis}
\label{sec:interpretability}

To support clinical interpretability and to examine consistency with ablation findings, we analyze model explanations using SHAP \cite{lundberg2017shap}. SHAP values are computed for tree-based models to estimate feature-level contributions to individual predictions.

Interpretability analyses are performed on representative test splits and focus on relative feature importance rather than absolute effect sizes. We use SHAP to assess whether the physiological signals identified as important through ablation exhibit corresponding attribution patterns at the model level. 

SHAP results are reported qualitatively and are used to corroborate trends observed in the ablation study, rather than as standalone evidence of causal relationships.

\section{Results}
\label{sec:results}

\subsection{Baseline Performance}

We evaluate baseline predictive performance under two regimes: a \emph{hospital-rich} setting using the full feature set (vitals, observations, laboratories, and notes), and an \emph{MCI-like} setting restricted to vitals only, simulating field triage conditions with limited sensing. 
Primary comparisons rely on threshold-independent metrics (AUROC and AUPRC) averaged over five patient-level splits, while representative-split figures are used to illustrate trends. Robustness across splits is further analyzed separately in section \ref{subsec:ablation}.

\paragraph{Hospital-rich features}
Left section of Table~\ref{tab:baseline_two_regimes_auc_ap} summarizes performance under the hospital-rich regime. Across models, tree-based ensemble methods consistently outperform linear baselines. XGBoost achieves the strongest overall discrimination, with a mean AUROC of $0.809$ and AUPRC of $0.382$, followed by Random Forest (AUROC $= 0.771$, AUPRC $= 0.332$). LightGBM attains competitive AUROC but exhibits weaker precision--recall performance, while Logistic Regression lags across both metrics. These results indicate that non-linear models are better able to exploit interactions among heterogeneous hospital signals, including early laboratory measurements and structured observations.

\paragraph{MCI-like features}
Right section of table~\ref{tab:baseline_two_regimes_auc_ap} reports performance when restricting inputs to vitals only. Despite the reduced feature set, performance degradation is modest. XGBoost again achieves the strongest ranking quality (AUROC $= 0.750$, AUPRC $= 0.296$), with Random Forest performing comparably (AUROC $= 0.755$, AUPRC $= 0.294$). Importantly, the relative ordering of models remains stable across regimes, suggesting that a substantial fraction of early deterioration signal is captured by basic physiological measurements alone. This observation supports the feasibility of early triage decision-support models  operating under limited sensing conditions.


\begin{table}[t]
\centering
\caption{Benchmark performance across feature regimes on the 10k cohort (mean $\pm$ SD over 5 patient-level splits). Hospital-rich uses all available features; MCI-like uses vitals only (including bedside consciousness and oxygenation indicators). {\footnotesize\textit{Bold = best/regime/metric. LR = Logistic Regression; RF = Random Forest.}}}
\label{tab:baseline_two_regimes_auc_ap}
\setlength{\tabcolsep}{3pt}
\renewcommand{\arraystretch}{1.15}
\begin{tabular}{lcccc}
\toprule
& \multicolumn{2}{c}{\textbf{Hospital-rich}} & \multicolumn{2}{c}{\textbf{MCI-like}} \\
\cmidrule(lr){2-3}\cmidrule(lr){4-5}
\textbf{Model} & \textbf{AUROC} & \textbf{AUPRC} & \textbf{AUROC} & \textbf{AUPRC} \\
\midrule
LR      & 0.668 ± 0.020 & 0.197 ± 0.012 & 0.704 ± 0.032 & 0.250 ± 0.042 \\
LGBM    & 0.777 ± 0.013 & 0.358 ± 0.016 & 0.692 ± 0.015 & 0.263 ± 0.022 \\
XGBoost & \textbf{0.809 ± 0.011} & \textbf{0.382 ± 0.026} & 0.750 ± 0.011 & \textbf{0.296 ± 0.022} \\
RF      & 0.771 ± 0.020 & 0.332 ± 0.037 & \textbf{0.755 ± 0.010} & 0.294 ± 0.018 \\
TabNet  & 0.786 ± 0.026 & 0.334 ± 0.035 & 0.744 ± 0.010 & 0.271 ± 0.023 \\
\bottomrule
\end{tabular}
\end{table}

\begin{figure}[t] 
        \centering
        \includegraphics[width=\linewidth]{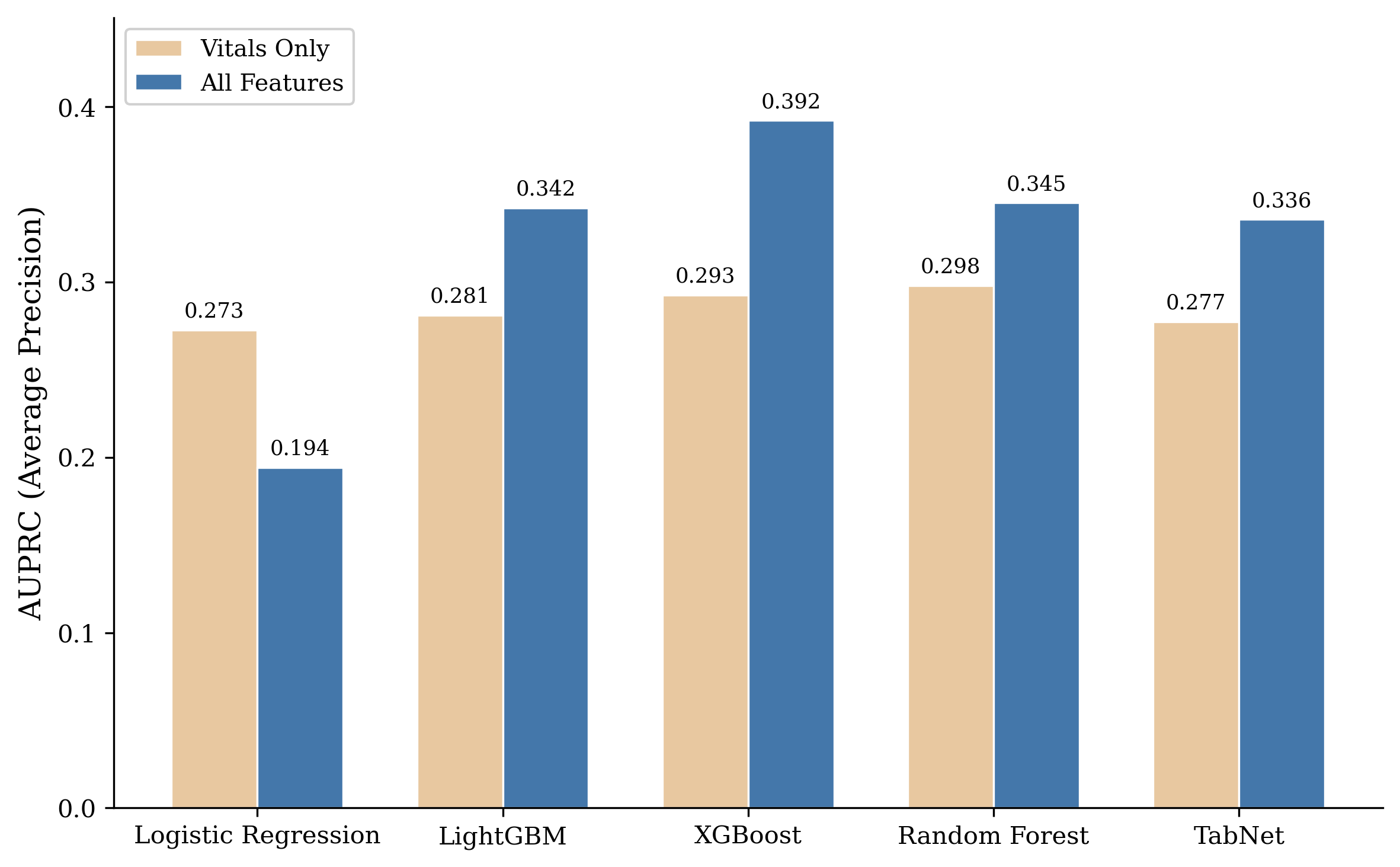}
        \caption{Average precision across models under MCI-like (vitals-only, including bedside consciousness, oxygenation) and hospital-rich (all-features) regimes. Results shown for a representative split.}
        \label{fig:perf-feature-groups}
    \end{figure}

\textit{Comparison across regimes and feature availability.}
Figure~\ref{fig:perf-feature-groups} compares representative-split AUPRC across models under hospital-rich (all-features) and MCI-like (vitals-only) regimes. Restricting inputs to vitals leads to a consistent reduction in average precision for all tree-based models, with XGBoost exhibiting the largest drop ($\downarrow 0.099$), followed by Random Forest ($\downarrow 0.047$). In contrast, Logistic Regression does not benefit from the addition of richer features and slightly degrades when moving to the full feature set, likely due to increased feature heterogeneity and multicollinearity.

    \begin{figure}[t] 
        \centering
        \includegraphics[width=\linewidth]{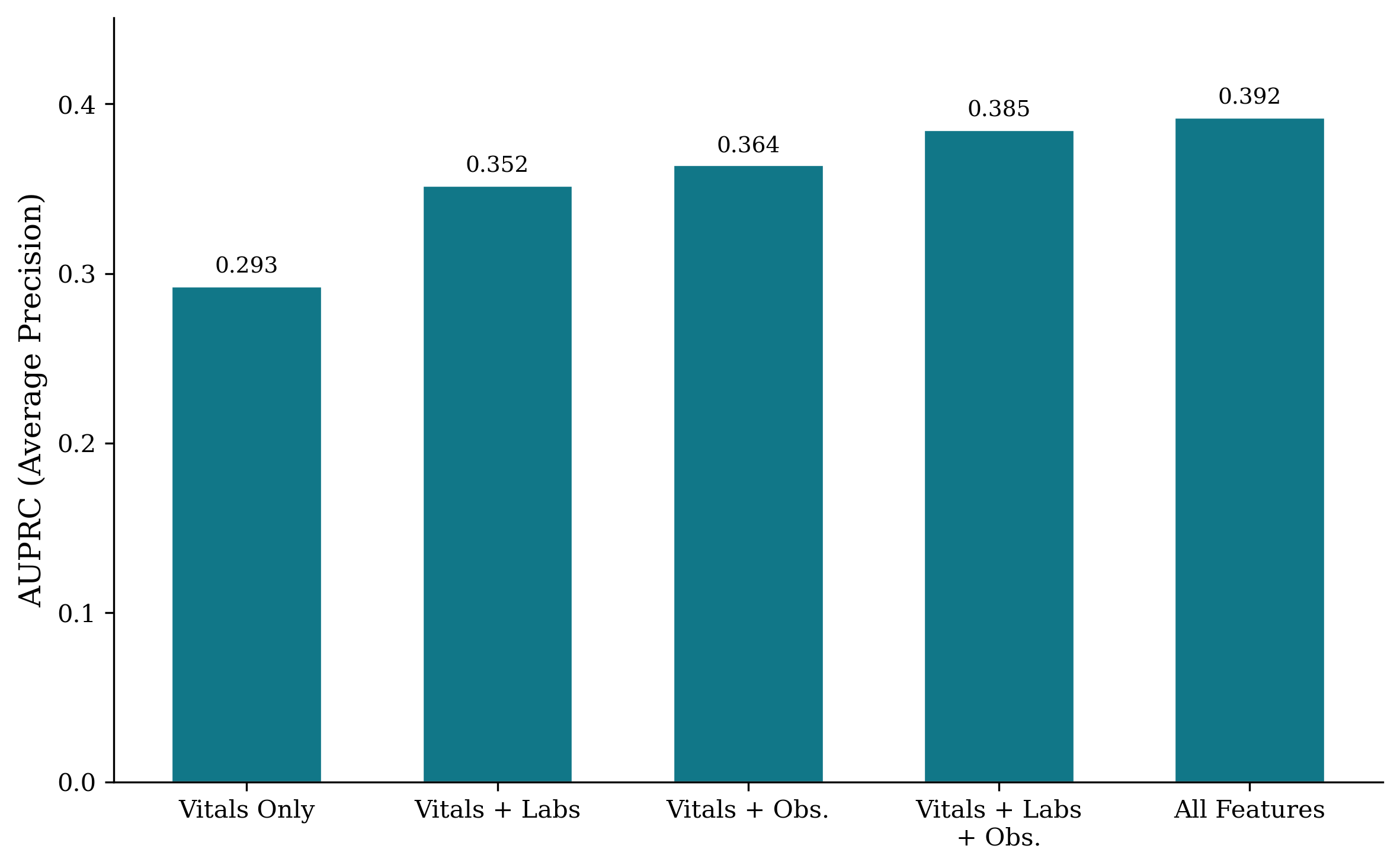}
        \caption{Effect of cumulative feature availability on Average Precision for XGBoost (representative split).}
        \label{fig:perf-cumulative}
    \end{figure}
    \label{fig:perf-sidebyside}

Figure~\ref{fig:perf-cumulative} further isolates this effect for XGBoost, showing that the dominant performance gain arises from the inclusion of observations beyond vitals ($\uparrow 0.071$), while laboratory features provide smaller but consistent incremental improvements. 
Across both analyses, changes are more pronounced in AUPRC than AUROC, consistent with the greater sensitivity of precision–recall metrics under class imbalance when contextual hospital information is unavailable. 
%
Together, these results indicate that while basic physiological measurements capture substantial early deterioration signal, observations provide the most impactful augmentation in field-constrained settings.





\textit{Operational performance.}
At a validated operating point on a representative patient-level split, ensemble models achieve F1-scores in the range of approximately $0.38–0.43$ and accuracies around $0.80–0.85$, depending on feature availability and model choice. Thresholds were selected separately for each feature regime using a held-out validation set on the representative split.
As these threshold-dependent metrics vary with operating point selection, primary conclusions throughout this section rely on threshold-independent measures (AUROC and AUPRC).

Overall, these results demonstrate that field-deployable, vitals-based models can retain substantial discriminative power, while modern ensemble methods provide robust and consistent performance across both hospital-rich and MCI-like triage settings.


\subsection{Ablation Study}
\label{subsec:ablation}
To evaluate robustness under partial observability typical of field and MCI-like triage, we analyze systematic vital-sign ablations on the 10k-patient cohort by progressively removing features. All models are retrained and evaluated under identical patient-level splits for each ablation condition, ensuring that observed performance differences reflect sensitivity to missing physiological signals rather than resampling variance.

The ablation results in Tables~\ref{tab:ablation-auroc} and~\ref{tab:ablation-auprc} reveal a consistent and interpretable ordering of vital sign importance across models and evaluation metrics. While models trained on all available features achieve the highest overall performance, vitals-only configurations retain substantial predictive power, indicating that early physiological measurements capture much of the information relevant to deterioration risk.
Leave-one-out ablations show that removing respiratory rate (RR) and temperature results in the largest degradations in performance, followed closely by oxygen saturation ($SpO_2$) and systolic blood pressure ($SBP$), whereas heart rate (HR) contributes comparatively less when considered in isolation. This pattern is observed consistently across both AUROC and AUPRC, highlighting the dominant role of respiratory and oxygenation-related signals in early triage prediction.
Across all ablation settings, performance trends are highly stable across random seeds, with standard deviations remaining small relative to effect sizes. XGBoost consistently achieve the strongest discrimination across both metrics, reinforcing the baseline findings that non-linear ensemble methods are best suited for early triage prediction.
Reduced vital subsets further demonstrate robustness under partial observability: configurations retaining respiratory or oxygenation signals consistently outperform those relying primarily on cardiovascular measures alone. 
%
Overall, the ablation study highlights the robustness of triage prediction under incomplete vital availability and identifies respiratory and oxygenation measures as the most critical signals in resource-limited or field-triage settings.





\begin{table*}[h]
\centering
\caption{Ablation study: AUROC (mean $\pm$ SD over 5 patient-level splits). Bold = best per regime.}
\label{tab:ablation-auroc}
\normalsize
\setlength{\tabcolsep}{8pt}
\begin{tabular}{@{}lccccc@{}}
\toprule
\textbf{Features} & \textbf{SGD} & \textbf{LGBM} & \textbf{XGB} & \textbf{TabNet} & \textbf{RF} \\
\midrule
All Features       & .668$\pm$.020 & .777$\pm$.013 & \textbf{.809$\pm$.011} & .786$\pm$.026 & .771$\pm$.020 \\
\midrule
Vitals Only        & .704$\pm$.032 & .692$\pm$.015 & .750$\pm$.011 & .744$\pm$.010 & \textbf{.755$\pm$.010} \\
$-$RR              & .535$\pm$.019 & .553$\pm$.015 & \textbf{.586$\pm$.013} & .510$\pm$.046 & .570$\pm$.009 \\
$-$SBP             & .527$\pm$.060 & .545$\pm$.012 & \textbf{.588$\pm$.014} & .568$\pm$.036 & .577$\pm$.014 \\
$-$HR              & .556$\pm$.024 & .557$\pm$.016 & \textbf{.599$\pm$.012} & .586$\pm$.038 & .583$\pm$.005 \\
$-$SpO$_2$         & .523$\pm$.032 & .550$\pm$.017 & \textbf{.598$\pm$.016} & .575$\pm$.018 & .579$\pm$.009 \\
$-$Temp            & .534$\pm$.042 & .553$\pm$.017 & \textbf{.591$\pm$.019} & .579$\pm$.012 & .576$\pm$.016 \\
\midrule
RR+HR+SBP          & .500$\pm$.038 & .552$\pm$.012 & \textbf{.590$\pm$.018} & .573$\pm$.011 & .576$\pm$.012 \\
RR+HR+Temp         & .533$\pm$.039 & .555$\pm$.016 & \textbf{.583$\pm$.018} & .567$\pm$.021 & .574$\pm$.009 \\
RR+Temp            & .547$\pm$.030 & .561$\pm$.021 & \textbf{.591$\pm$.009} & .580$\pm$.008 & .564$\pm$.013 \\
RR+HR              & .518$\pm$.053 & .561$\pm$.021 & \textbf{.577$\pm$.016} & .574$\pm$.017 & .563$\pm$.017 \\
RR only            & .543$\pm$.065 & .601$\pm$.011 & \textbf{.602$\pm$.011} & .584$\pm$.005 & .599$\pm$.013 \\
HR only            & .489$\pm$.043 & .534$\pm$.018 & \textbf{.542$\pm$.015} & .522$\pm$.030 & .533$\pm$.015 \\
\bottomrule
\end{tabular}
\end{table*}

\begin{table*}[h]
\centering
\caption{Ablation study: AUPRC (mean $\pm$ SD over 5 patient-level splits). Bold = best per regime.}
\label{tab:ablation-auprc}
\normalsize
\setlength{\tabcolsep}{8pt}
\begin{tabular}{@{}lccccc@{}}
\toprule
\textbf{Features} & \textbf{SGD} & \textbf{LGBM} & \textbf{XGB} & \textbf{TabNet} & \textbf{RF} \\
\midrule
All Features       & .197$\pm$.012 & .358$\pm$.016 & \textbf{.382$\pm$.026} & .334$\pm$.035 & .332$\pm$.037 \\
\midrule
Vitals Only        & .250$\pm$.042 & .263$\pm$.022 & \textbf{.296$\pm$.022} & .271$\pm$.023 & .294$\pm$.018 \\
$-$RR              & .154$\pm$.015 & .170$\pm$.012 & \textbf{.183$\pm$.006} & .135$\pm$.025 & .177$\pm$.009 \\
$-$SBP             & .148$\pm$.022 & .168$\pm$.008 & \textbf{.190$\pm$.013} & .174$\pm$.021 & .183$\pm$.007 \\
$-$HR              & .160$\pm$.019 & .182$\pm$.010 & \textbf{.192$\pm$.015} & .175$\pm$.019 & .187$\pm$.010 \\
$-$SpO$_2$         & .157$\pm$.007 & .173$\pm$.007 & \textbf{.191$\pm$.014} & .178$\pm$.009 & .185$\pm$.010 \\
$-$Temp            & .143$\pm$.017 & .158$\pm$.011 & \textbf{.171$\pm$.013} & .160$\pm$.008 & .170$\pm$.012 \\
\midrule
RR+HR+SBP          & .131$\pm$.018 & .158$\pm$.007 & \textbf{.171$\pm$.011} & .161$\pm$.005 & .170$\pm$.009 \\
RR+HR+Temp         & .159$\pm$.030 & .175$\pm$.007 & \textbf{.188$\pm$.012} & .166$\pm$.019 & .181$\pm$.007 \\
RR+Temp            & .166$\pm$.020 & .188$\pm$.016 & \textbf{.192$\pm$.012} & .177$\pm$.010 & .185$\pm$.008 \\
RR+HR              & .136$\pm$.022 & .161$\pm$.013 & \textbf{.170$\pm$.010} & .165$\pm$.009 & .165$\pm$.011 \\
RR only            & .146$\pm$.023 & .172$\pm$.007 & \textbf{.173$\pm$.008} & .154$\pm$.005 & .170$\pm$.008 \\
HR only            & .119$\pm$.013 & .151$\pm$.010 & \textbf{.154$\pm$.008} & .135$\pm$.015 & .152$\pm$.010 \\
\bottomrule
\end{tabular}
\end{table*}

\subsection{ICU Transfer as a Secondary Deterioration Outcome}
We additionally evaluate ICU transfer as an alternative deterioration endpoint to contrast
decision-making under acute escalation versus mortality. Unlike mortality, which reflects
broader systemic severity over the course of hospitalization, ICU transfer represents a
short-horizon escalation decision driven by immediate physiological instability.

Using XGBoost and the same patient-level splits as in prior experiments, ICU-transfer
prediction exhibits a smaller performance gap when restricted to vitals-only features.
Specifically, the AUPRC decreases from 0.815 with all available features to 0.719 under
vitals-only inputs, indicating that a substantial fraction of escalation-relevant signal is
already captured by first-hour physiological measurements.

To further characterize endpoint-specific sensitivity, we perform a leave-one-out ablation
over vital signs using the vitals-only model as the reference. Table~\ref{tab:loo_vitals_icu_xgb}
reports the exact AUPRC changes when each vital is removed individually. Removing systolic
blood pressure produces the largest degradation, followed by
temperature, respiratory rate , oxygen saturation, and
heart rate, in order. Notably, multiple respiratory and circulatory variables contribute
comparably, and no single vital alone dominates model performance.

These results indicate that ICU transfer prediction relies on a combination of acute
hemodynamic and respiratory instability, with greater redundancy across vitals compared
to mortality prediction. In contrast to mortality, where systolic blood pressure exhibits
a more pronounced marginal role, ICU escalation decisions appear more directly tied to
short-term physiological status that can be assessed using readily available vital signs.

This leave-one-out analysis is intended to characterize relative dependence on
individual physiological signals rather than statistical uncertainty; accordingly,
we report $\Delta$AUPRC as effect size; repeating this analysis across seeds did not change the observed ordering.

\begin{table}[h]
\centering
\caption{Leave-one-out analysis of vital signs for ICU deterioration prediction
(XGBoost, AUPRC). $\Delta$AUPRC is computed relative to the vitals-only baseline.}
\label{tab:loo_vitals_icu_xgb}
\begin{tabular}{lcc}
\toprule
\textbf{Feature set} & \textbf{AUPRC} & \textbf{$\Delta$AUPRC} \\
\midrule
All features & 0.815099 ~$\pm$~ 0.002 & -- \\
Vitals only & 0.718715 ~$\pm$~ 0.001 & 0.000 \\
\midrule
Vitals $-$ SBP  & 0.678296 & $-0.040$ \\
Vitals $-$ Temp & 0.692633 & $-0.026$ \\
Vitals $-$ RR   & 0.694714 & $-0.024$ \\
Vitals $-$ HR   & 0.700065 & $-0.019$ \\
Vitals $-$ SpO$_2$ & 0.700558 & $-0.018$ \\
\bottomrule
\end{tabular}
\end{table}

\subsection{Model Interpretability via SHAP}

\begin{figure*}[htbp]
\centering
\subfloat[Hospital-rich features]{%
\includegraphics[width=0.32\linewidth]{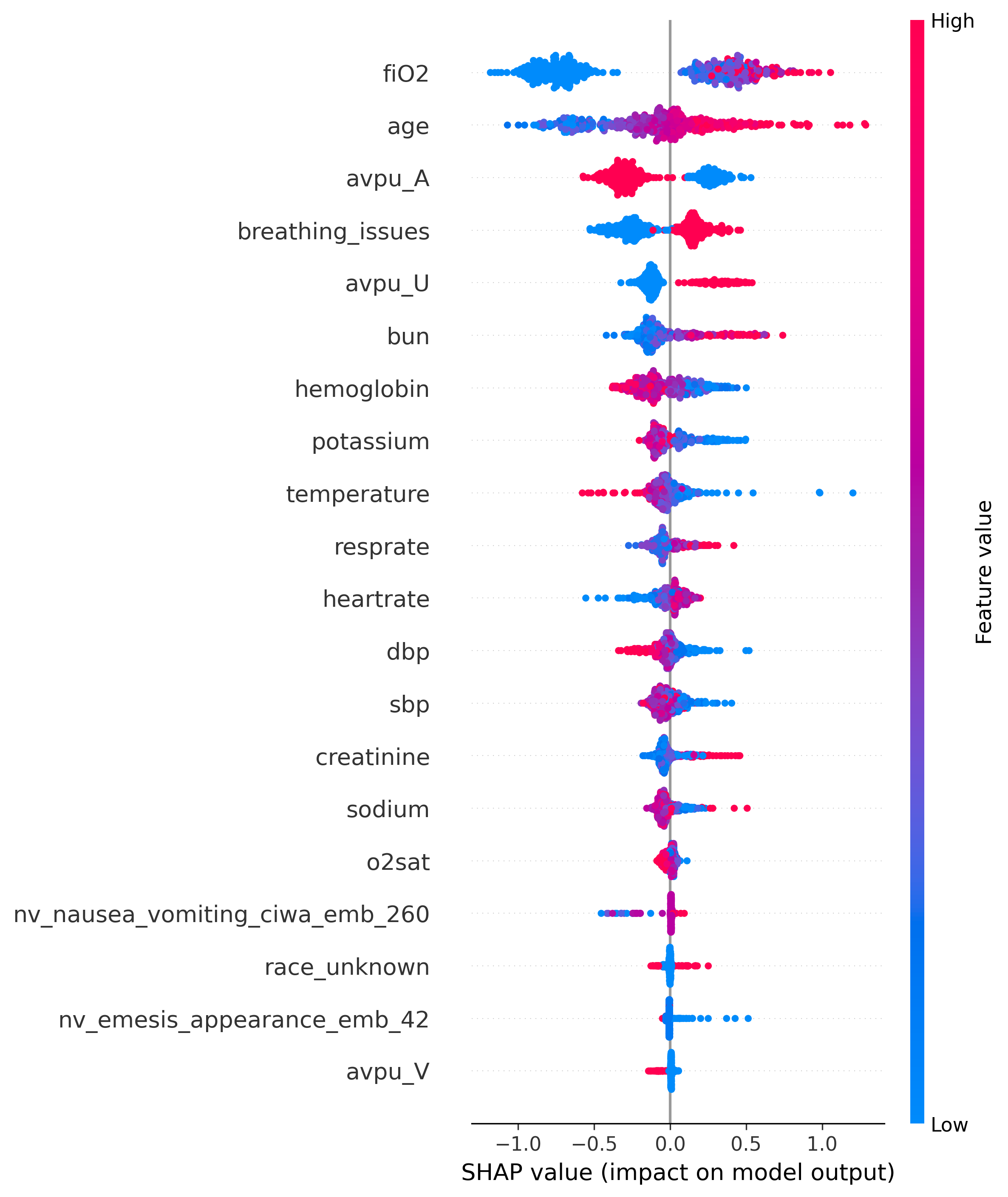}
\label{fig:shap-xg-full}
}
\subfloat[Labs and Vitals features]{%
\includegraphics[width=0.32\linewidth]{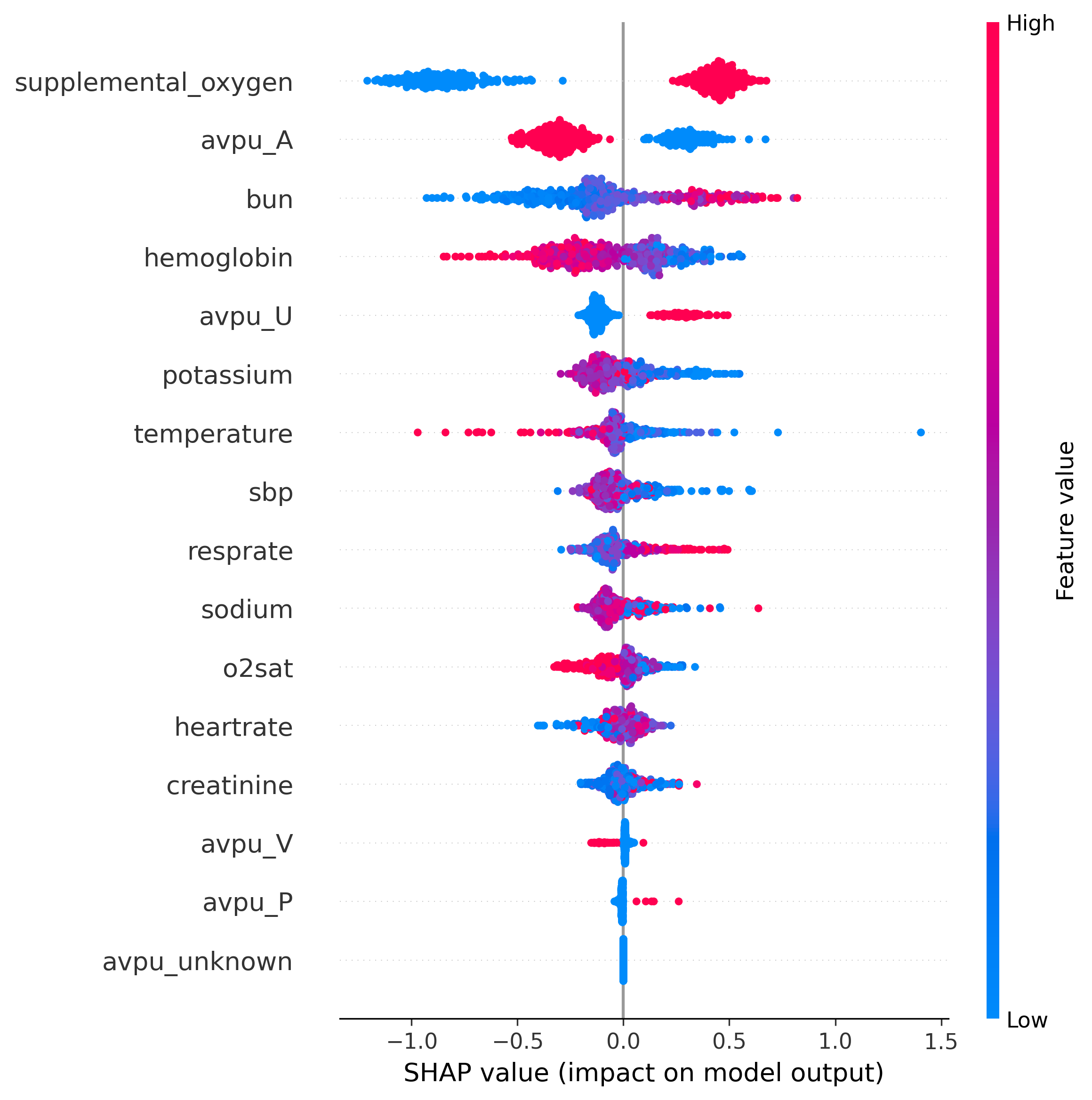}
\label{fig:shap-xg-V&L}
}
\subfloat[MCI-like features]{%
\includegraphics[width=0.32\linewidth]{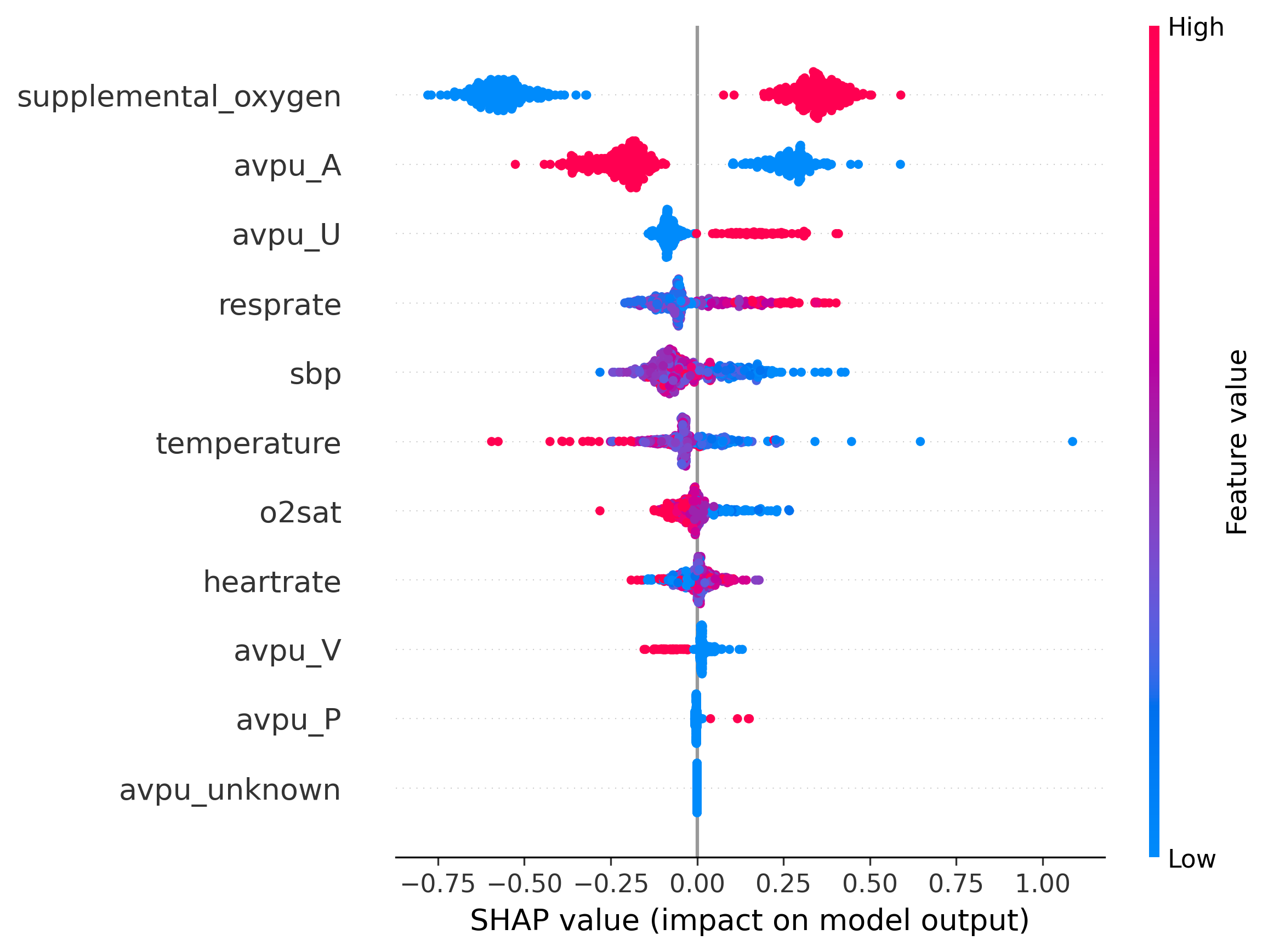}
\label{fig:shap-xg-mci}
}

\caption{Global SHAP summaries for the XGBoost model under three feature regimes:
(a) hospital-rich;
(b) vitals and laboratory;
(c) MCI-like setting using vitals only.
Each point corresponds to a patient, colored by feature value (red = high, blue = low),
with horizontal position indicating SHAP contribution to predicted deterioration risk.}
\label{fig:shap-comparison}
\end{figure*}

To explain how the best-performing models utilize available clinical information under
different resource conditions, we apply SHAP (Shapley Additive Explanations) to the
XGBoost models across three feature regimes (Fig.~\ref{fig:shap-comparison}):
(a) hospital-rich features, (b) labs and vitals, and (c) MCI-like reduced features.
%

Across all three regimes, attribution patterns are dominated by oxygenation- and
respiratory-related signals together with neurologic responsiveness. In particular,
supplemental oxygen use (and FiO$_2$ in the hospital-rich setting), oxygen saturation,
and respiratory rate consistently appear among the highest-ranked contributors
Neurologic responsiveness, as captured by AVPU, also emerges as a strong contributor, alongside hemodynamic measures such as systolic blood pressure. 
These patterns align closely with the ablation
results, which show the largest degradation 
when oxygen saturation, systolic blood pressure, or respiratory rate are removed, and
comparatively smaller effects when heart rate is removed in isolation.

In the hospital-rich regime (Fig.~\ref{fig:shap-xg-full}), laboratory variables such as
blood urea nitrogen, hemoglobin, potassium, and creatinine contribute measurable but
secondary signal alongside age and core vitals. While these laboratory features refine
risk estimates, their effects are smaller and less consistently directional than those
of oxygenation and respiratory variables, consistent with the modest and sometimes
variable performance gains observed when early laboratory data are added in the
benchmark and cumulative analyses.

When restricting inputs to labs and vitals (Fig.~\ref{fig:shap-xg-V&L}), the overall
structure of feature attribution remains largely unchanged. Oxygen support, respiratory
measures, and AVPU continue to dominate, while laboratory features provide supportive
but non-dominant contributions. This suggests that much of the discriminative signal for
early deterioration is already captured by immediately observable physiological
measurements.

In the MCI-like regime (Fig.~\ref{fig:shap-xg-mci}), attributions become more compact and
field-realistic. Supplemental oxygen use, AVPU, respiratory rate, systolic blood
pressure, temperature, and oxygen saturation account for the majority of the predictive
signal, indicating that the core structure of model reasoning is preserved even under
substantially reduced feature availability. This observation is consistent with the
relatively modest performance degradation observed in the vitals-only setting and
supports the feasibility of early triage decision-support models  operating under limited
sensing conditions.

Finally, note-derived embedding components appear only at low rank in the
hospital-rich regime and do not emerge as dominant contributors. This suggests that
unstructured symptom text provides at most auxiliary benefit relative to immediately
observable physiological and observational signals during early triage.

\section{Discussion and Future Work}
This work demonstrates that meaningful early deterioration prediction is feasible using only information available during the initial ED assessment, even under feature-limited, MCI-like conditions. 
Across models and outcomes, basic physiological measurements retain substantial discriminative power, with performance degradation from hospital-rich to MCI-like regimes remaining modest. This finding supports the clinical intuition that early instability is often reflected in immediately observable vital signs rather than downstream laboratory confirmation.
The ablation and interpretability analyses consistently identify oxygenation and respiratory measures as the most influential contributors, followed by blood pressure and temperature, while heart rate alone provides weaker marginal signal. The stability of this ordering across metrics, models, and random splits suggests that these patterns reflect underlying physiological relevance rather than model-specific artifacts. Moreover, reduced vital subsets preserve non-trivial performance, indicating redundancy among measurements and robustness to partial sensor loss -- an important property for field or resource-constrained triage settings.


Several limitations and extensions remain. This study focuses on single-encounter, snapshot-based prediction; extending the framework to longitudinal trajectories and dynamic updates over the ED course is a natural next step. Future work will also explore the incorporation of knowledge-infused proxy features and systematic benchmarking against deep sequence models such as RETAIN and AcuityNet, under controlled partial-observability settings. While we evaluate a large, curated 10k-patient cohort derived from MIMIC-IV-ED, full-scale benchmarking on the complete dataset and validation on external ED or prehospital cohorts will be necessary to assess generalizability. Finally, future work may explore more realistic simulations of MCI scenarios, including structured sensor loss or delayed measurements, to further stress-test robustness under extreme partial observability.



\section{Conclusion}
We presented a leakage-aware, reproducible benchmark for early deterioration prediction at emergency department triage, explicitly connecting hospital-rich modeling with field-constrained, MCI-like settings. Using only information available within the first hour of ED arrival and enforcing patient-level independence, we showed that basic physiological measurements retain substantial predictive power, even under restricted sensing. Across models, non-linear ensemble methods consistently achieved the strongest performance, while ablation and interpretability analyses identified oxygenation and respiratory signals as the most critical contributors to early risk. 
By quantifying the performance gap between hospital-rich and vitals-based regimes and demonstrating resilience to partial vital availability, this work highlights the feasibility of deployable triage models under real-world constraints. We hope this benchmark and analysis framework will facilitate more transparent, comparable, and clinically grounded research on early triage prediction across diverse care settings.

\bibliographystyle{plain}
\bibliography{refs}

@inproceedings{chang2024edtriage,
  title={Deep learning to predict hospitalization at triage: Integration of structured data and unstructured text},
  author={Arnaud, {\'E}milien and Elbattah, Mahmoud and Gignon, Maxime and Dequen, Gilles},
  booktitle={2020 IEEE International Conference on Big Data (Big Data)},
  pages={4836--4841},
  year={2020},
  organization={IEEE}
}

@article{start1996,
  title={Current Classifications and Triage Scoring Scales.},
  author={Straus, Slavenka},
  journal={Journal of Anesthesia/Anestezi Dergisi (JARSS)},
  volume={33},
  year={2025}
}

@article{johnson2023mimiciv,
  title        = {MIMIC-IV, a freely accessible electronic health record dataset},
  author       = {Johnson, Alistair E.\ W. and Bulgarelli, Lucas and Shen, Lu and Gayles, Alvin and Shammout, Ayad and Horng, Steven and Pollard, Tom J. and Hao, Sicheng and Moody, Benjamin and Gow, Brian and Lehman, Li-Wei H. and Celi, Leo A. and Mark, Roger G.},
  journal      = {Scientific Data},
  volume       = {10},
  number       = {1},
  pages        = {1},
  year         = {2023},
  doi          = {10.1038/s41597-022-01899-x},
}

@article{salt2008,
  title={Modern triage in the emergency department},
  author={Christ, Michael and Grossmann, Florian and Winter, Daniela and Bingisser, Roland and Platz, Elke},
  journal={Deutsches {\"A}rzteblatt International},
  volume={107},
  number={50},
  pages={892},
  year={2010}
}

@article{liu2021prospective,
  title={Artificial Intelligence-Assisted Triage Accuracy in Pediatric Emergency Departments Across East Mediterranean Hospitals: A Multicenter Validation Study: Jessica Taylor, Michael Chen, Aisha Williams, Daniel Reed, Emily Parker},
  author={Taylor, Jessica and Chen, Michael and Williams, Aisha and Reed, Daniel and Parker, Emily},
  journal={Ambulatory Pediatrics},
  volume={9},
  number={07},
  pages={155--167},
  year={2025}
}

@article{liu2025gradient,
  title={Using machine learning and natural language processing in triage for prediction of clinical disposition in the emergency department},
  author={Chang, Yu-Hsin and Lin, Ying-Chen and Huang, Fen-Wei and Chen, Dar-Min and Chung, Yu-Ting and Chen, Wei-Kung and Wang, Charles CN},
  journal={BMC Emergency Medicine},
  volume={24},
  number={1},
  pages={237},
  year={2024},
  publisher={Springer}
}

@article{lundberg2017shap,
  title={A unified approach to interpreting model predictions},
  author={Lundberg, Scott M and Lee, Su-In},
  journal={Advances in neural information processing systems},
  volume={30},
  year={2017}
}

@article{roberts2023mimiced,
  title={Benchmarking emergency department prediction models with machine learning and public electronic health records},
  author={Feng Xie et al.},
  journal={Scientific Data},
  volume={9},
  number={1},
  pages={658},
  year={2022},
  publisher={Nature Publishing Group UK London}
}

@article{sacco2005saves,
  title={Operational comparison of the Simple Triage and Rapid Treatment method and the Sacco Triage Method in mass casualty exercises},
  author={Navin, D Michael and Sacco, William J and Waddell, Robert},
  journal={Journal of Trauma and Acute Care Surgery},
  volume={69},
  number={1},
  pages={215--225},
  year={2010},
  publisher={LWW}
}

@article{sitthiprawat2025electronic,
  title={Development and internal validation of an AI-based emergency triage model for predicting critical outcomes in emergency department},
  author={Sitthiprawiat, Patipan and Wittayachamnankul, Borwon and Sirikul, Wachiranun and Laohavisudhi, Korsin},
  journal={Scientific Reports},
  volume={15},
  number={1},
  pages={31212},
  year={2025},
  publisher={Nature Publishing Group UK London}
}

@article{bmj2021news2,
  title={Evaluation of the utility of NEWS2 during the COVID-19 pandemic},
  author={Williams, Bryan},
  journal={Clinical Medicine},
  volume={22},
  number={6},
  pages={539--543},
  year={2022},
  publisher={Elsevier}
}

@article{royal2017news2,
  title={The national early warning score 2 (NEWS2)},
  author={Smith, Gary B and Redfern, Oliver C and Pimentel, Marco AF and Gerry, Stephen and Collins, Gary S and Malycha, James and Prytherch, David and Schmidt, Paul E and Watkinson, Peter J},
  journal={Clinical Medicine},
  volume={19},
  number={3},
  pages={260},
  year={2019}
}

@article{xu2023mci,
  title={Triage in major incidents: development and external validation of novel machine learning-derived primary and secondary triage tools},
  author={Yuanwei Xu et al.},
  journal={Emergency Medicine Journal},
  volume={41},
  number={3},
  pages={176--183},
  year={2024},
  publisher={BMJ Publishing Group Ltd and the British Association for Accident~…}
}

@article{boulitsakis2023predicting,
  title={Predicting acute clinical deterioration with interpretable machine learning to support emergency care decision making},
  author={Boulitsakis Logothetis, Stelios and Green, Darren and Holland, Mark and Al Moubayed, Noura},
  journal={Scientific reports},
  volume={13},
  number={1},
  pages={13563},
  year={2023},
  publisher={Nature Publishing Group UK London}
}

@misc{johnson2024mimic,
  title={MIMIC-IV (version 3.0). PhysioNet},
  author={Johnson, A and Bulgarelli, L and Pollard, T and Gow, B and Moody, B and Horng, S and Celi, LA and Mark, R},
  year={2024}
}

@article{arik2021tabnet,
  title={TabNet: Attentive Interpretable Tabular Learning},
  author={Arik, Sercan O and Pfister, Tomas},
  journal={Proceedings of the AAAI Conference on Artificial Intelligence},
  year={2021}
}

@misc{pubmedbert,
  author = {Yu Gu et al.},
  title = {Domain-Specific Language Model Pretraining for Biomedical Natural Language Processing},
  year = {2020},
  eprint = {arXiv:2007.15779},
}

@article{PhysioNet-mietic-1.0.0,
  author = {Shen, Qingyang and Guo, Quan},
  title = {{MIMIC-IV-Ext Triage Instruction Corpus}},
  journal = {{PhysioNet}},
  year = {2025},
  month = mar,
  note = {Version 1.0.0},
  doi = {10.13026/q1nc-2e47},
  url = {https://doi.org/10.13026/q1nc-2e47}
}

@article{erwander2025vitals,
  title={The role of vital signs in predicting mortality risk in elderly patients visiting the emergency department},
  author={Erwander, Karin and Agvall, Bj{\"o}rn and Ivarsson, Kjell},
  journal={BMC Emergency Medicine},
  year={2025},
  doi={10.1186/s12873-025-01307-8},
  pmid={40750848}
}

@article{lin2024interpretable,
  title={Interpretable Deep Learning System for Identifying Critical Patients Through the Prediction of Triage Level, Hospitalization, and Length of Stay},
  author={Yu-Ting Lin et al},
  journal={JMIR Medical Informatics},
  year={2024},
  pmid={38557661}
}

\end{document}